\documentclass[a4paper]{jpconf}
\usepackage{graphicx}
\usepackage{color}
\usepackage{iopams}

\begin{document}
\title{Electron Correlation and Spin Dynamics in Iron Pnictides and Chalcogenides}

\author{Rong Yu$^{1}$, Qimiao Si$^{1}$, Pallab Goswami$^{2}$, Elihu Abrahams$^{3}$}

\address{$^{1}$ Department of Physics and Astronomy, Rice University,
Houston, Texas 77005, USA}
\address{$^{2}$ National High Magnetic Field Laboratory and Department of Physics,
Florida State University, Tallahassee, Florida 32306, USA}
\address{$^{3}$ Department of Physics and Astronomy,
University of California Los Angeles, Los Angeles, CA 90095, USA}

\ead{qmsi@rice.edu}

\begin{abstract}
Superconductivity in the iron pnictides and chalcogenides is closely connected to a bad-metal
normal state and a nearby antiferromagnetic order. Therefore,
considerable attention has been focused on the role of electron correlations and spin dynamics.
In this article, we summarize some key experiments that quite directly imply strong
electron correlations in these materials, and discuss  aspects of the recent theoretical studies
on these issues.  In particular, we outline a $w$-expansion, which treats the correlation effects
using the Mott transition as the reference point.
For the parent systems, it gives rise to an effective $J_1$-$J_2$ model
that is coupled to the itinerant electrons in the vicinity of the Fermi energy; this model
yields an isoelectronically-tuned quantum critical point, and allows a study of the distribution
of the spin spectral weight in the energy and momentum space in the paramagnetic phase.
Within the same framework, we demonstrate the Mott insulating phase in the iron oxychalcogenides
as well as the alkaline iron selenides; for the latter system, we also consider the role
of an orbital-selective Mott phase. Finally,  we discuss the singlet superconducting pairing driven
by the short-range $J_1$-$J_2$ interactions. Our considerations highlight the iron pnictides
and chalcogenides as exemplifying  strongly-correlated electron systems at the boundary of
electronic localization and itinerancy.
\end{abstract}

\section{Introduction}

Superconductivity in iron pnictides and chalcogenides \cite{Hosono}
occurs by chemical
doping of their antiferromagnetic parents, compounds in which
the Fe valence is $+2$.
The relationship between superconductivity and parent antiferromagnetism
has been discussed from the beginning of the field \cite{Cruz}.
The materials have a layered structure, with each layer containing
a square lattice of Fe ions. Their electronic properties are primarily
associated with the Fe 3$d$ orbitals. The appropriate microscopic
Hamiltonian contains a kinetic-energy part involving both intra-
and inter-orbital hopping, as well as an interaction component
featuring Coulomb repulsions and Hund's coupling.

The study of spin dynamics has played a central role in the developing understanding of the microscopic physics 
and the phases of these compounds. Based on the fact that the parent compounds are
``bad" metals, a strong-coupling approach was advanced early on \cite{SiAbrahams}.
This approach expands around the Mott transition in terms of a parameter $w$,
which measures the coherent part of electronic excitation spectrum as is depicted in Fig.\ 1;
this parameter
is small when the ratio of effective electron-electron interaction to bandwidth is large.
\begin{figure}[h]
\centering
\includegraphics[totalheight=0.3\textheight, viewport=-50 120 800 480,clip]{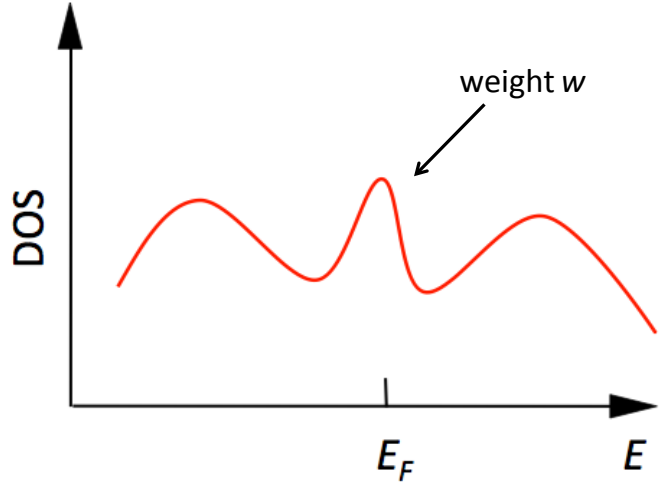}%
\vskip -1pc
\caption{\label{fig1}{\bf Single-electron spectral function as the sum of
coherent and incoherent parts.}
The single-electron density
of states (DOS) is plotted against energy ($E$); $E_F$ is the
Fermi energy.
Each peak may
contain additional structure
due to the multi-orbital nature of the iron
pnictides.
 }
\end{figure}

To zeroth order  in $w$, the entire spectrum is incoherent and represents quasi-localized local moments
and a $J_1$-$J_2$ coupling of these local moments was proposed \cite{SiAbrahams,AbInitio};
this provided the understanding of the collinear antiferromagnetic order of the parent compounds.
At higher order in $w$, the local moments become coupled to the itinerant carriers \cite{SiAbrahams,SiNJP,DaiPNAS}. 
In the doped systems, the approach leads to  a multiband
$t$-$J_1$-$J_2$ model that has been used to study the superconductivity. This is a strong-coupling
approach, and is in sharp contrast to a weak-coupling description that invokes nesting of hole
and electron Fermi surface pockets to drive both the spin fluctuations and superconductivity.
Related strong-coupling approaches have been taken from a number of
perspectives
\cite{Yin,KSeo,WChen,Moreo,Berg,Lv,Ma,MJHan,Wysocki,Fang:08,Xu:08,Uhrig,
Ishida,Lorenzana,Bascones}.

Here we provide a brief status report on this line of approach to the iron pnictides.
We begin by expounding, in Sec.~\ref{bad_metal},  on the experimental evidences
for the importance of electronic correlations.

In Sec.~\ref{w_expansion}, we summarize the $w$-expansion with respect to the Mott transition point.
Sec.~\ref{qcp} is devoted to the magnetic quantum critical point (QCP) that arises by increasing $w$,
which is associated with the reduction of $U/W$, the ratio of the direct Coulomb and Hund's exchange
interactions to the characteristic bandwidth. In addition, we summarize the
 theoretical proposal for realizing the QCP by P-substitution for As in the parent
iron pnictides, and the subsequent experimental realizations.

In Sec.~\ref{spin_dynamics}, we consider the
quantum fluctuations in the magnetic sector. For the parent iron pnictides, we show that our
$J_1$-$J_2$ based results compare
well with the inelastic neutron scattering results in the parent iron pnictides. For the parent
alkaline
iron selenides, the predictions based on an extended $J_1$-$J_2$ model have been confirmed by
experiments.

In Sec.~\ref{mott_oxyselenides}, we  consider the effects of
tuning the parent systems in the opposite direction,
by enhancing $U/W$. This occurs in La$_2$O$_2$Fe$_2$O(Se,S)$_2$,
in which the expansion of the Fe square lattice leads to
a Mott insulator.

In Sec.~\ref{kfese}, we 
address the correlation effects in the alkaline iron selenides.
The enhanced electron correlations by the ordered iron vacancies leads to
Mott insulating behavior in the parent compounds. We  
consider 
how the vacancy-ordered Mott insulating parent compounds are connected
to the vacancy-disordered metallic phase via a novel orbital-selective Mott phase.
Finally, we discuss the effect of the reduced/suppressed vacancy order 
on the spin spectral weight at temperatures above the N\'eel/structural transitions 
in the insulating alkaline iron selenides.

In Sec.~\ref{sc}, we discuss the superconducting pairing for iron pnictides and alkaline
iron selenides
based on the strong-coupling model. Sec.~\ref{summary} includes some summary remarks.

\begin{figure}[t!]
\centering
\includegraphics[totalheight=0.50\textheight, viewport=-70 10 600 520,clip]{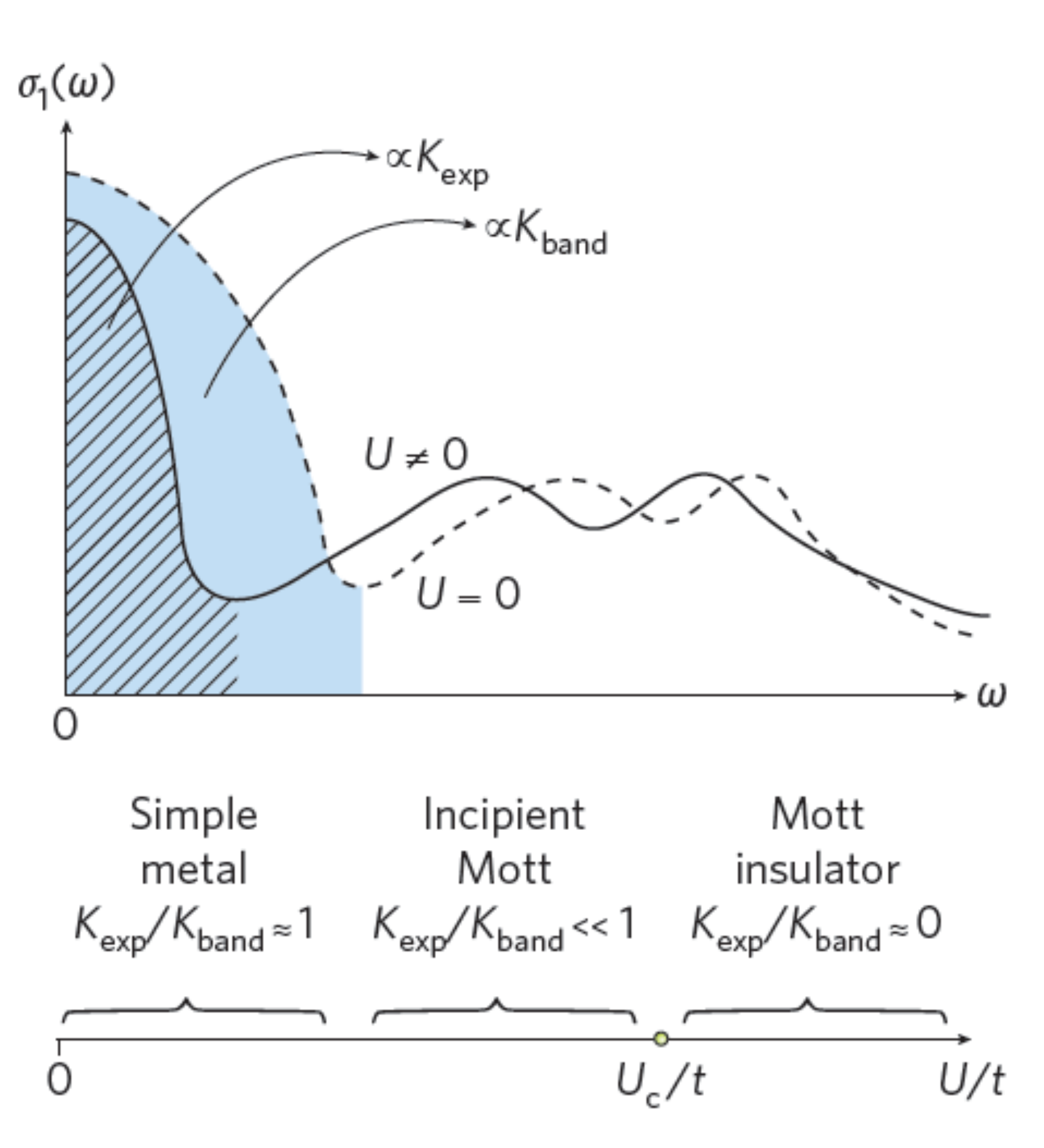}
\vskip -0.2pc
\caption{\label{QMCap}
{\bf Electronic correlations in the undoped
iron pnictides.} The optical conductivity as a function
of frequency. The area $K$ under the $\omega = 0$ Drude
peak is
proportional to the kinetic energy of the coherent
electrons near the Fermi energy. $K_{exp}$ is measured and $K_{band}$, is the expected value
from (non-interacting) band theory.
The degree to which $K_{exp} /K_{band}$ is smaller than one is a measure of the ratio
of the electron-electron repulsion to bandwidth, $U/t$, or a measure
of the coherent quasiparticle weight $w$.
 The interaction transfers spectral weight from low energies to high energies, up to order $U$.
 $U_c$ is the threshold interaction for a Mott localization transition. That $K_{exp} /K_{band}$
 is substantially smaller than one means that $U/t$ is smaller than but close to $U_c /t$,
 where the electrons are on the verge of losing their itinerancy.}
\end{figure}

\section{Experimental manifestations of strong-coupling physics}
\label{bad_metal}
The strong-coupling approach has its roots in a number of experimental observations.
One is the fact that the room temperature resistivity is so high ($\sim 0.4 \;$ m$\Omega$-cm) that the
mean free path is only of order the Fermi wavelength, as is typical for bad metals near a Mott transition. Because this is the case
even in compounds that are quite pure as inferred from their small residual resistivity, it signifies
the existence of strong inelastic scatterings that are attributable to the sizable electron-electron
interactions \cite{SiAbrahams,AbrahamsSi2011}. This arises if the metallic parent iron pnictides
are close to a Mott transition, when only a small
portion of the single-electron excitation spectrum lies in the coherent low-energy part,
of weight $w$ (Fig.\ 1).

This bad-metal interpretation has been confirmed by optical conductivity measurements
\cite{Qazilbash,HuOptics,YangOptics,Boris,Degiorgi}.
As illustrated in Fig.\ 2, the experimentally measured spectral weight for the Drude peak
normalized by the non-interacting value extracted from band-structure calculations,
$K_{exp}/K_{band}$,  provides a measure of the correlation strength: it would be close to $1$
for weakly interacting systems, and close to $0$ for Mott insulators.
The experimentally determined
value \cite{Qazilbash,HuOptics} is close to about $1/3$, falling into the regime of bad metals
in proximity to Mott localization. This implies that a large portion of the single-electron
spectral weight lies in the incoherent part,
which would be associated with the large Coulomb (direct plus Hund's) interaction energy scale.
At the same time, this reduction of Drude weight is accompanied by the transfer
of optical spectral weight between low and high frequencies.

\begin{figure}[t!]
\centering
\includegraphics[width=28pc]{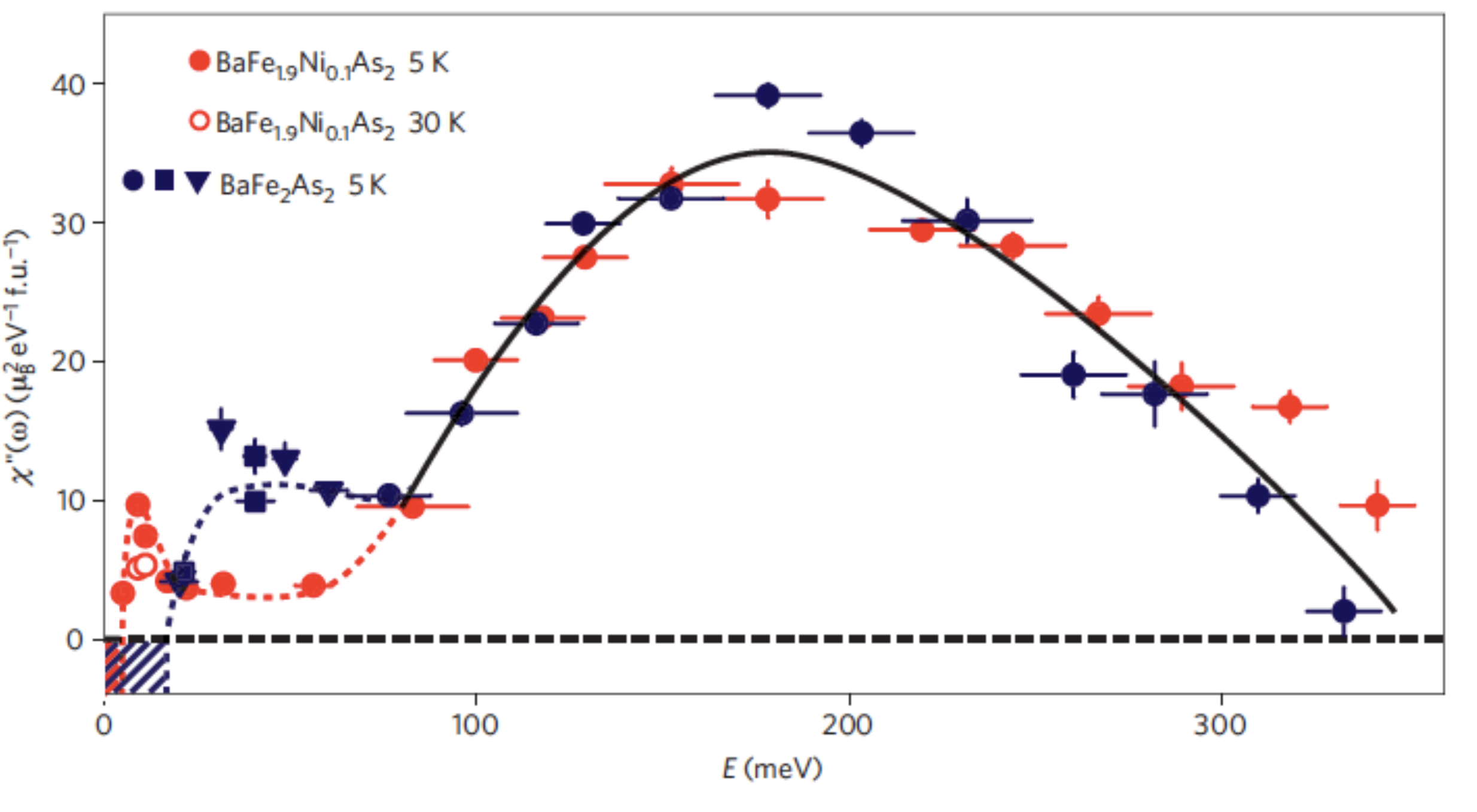}
\caption{\label{fig3}{\bf Spin spectral weight for several iron arsenides.} Adapted with permission from Ref.~\cite{LiuNatPhys12}.}
\end{figure}

The spin dynamics, as revealed in inelastic neutron scattering
experiments \cite{LiuNatPhys12, Zaliznyak11,ZhaoNatPhys09},
also gives strong support to the picture
of interacting local moments, whose sizeable spectral weight arises from the
{\it incoherent} part  (weight $\sim 1-w$) of the electronic excitation spectrum.
This is illustrated in Fig.\ 3, adapted from Liu et al \cite{LiuNatPhys12}, where the spin
spectral distribution is shown over a range of energies which includes spin waves throughout
the Brillouin zone. The integrated spectral weight is very large
($\int d\omega \chi'' \sim 3 \mu_B^2$ per Fe), consistent with local moments of spin $\sim$ 1.
Such a large spectral weight readily arises
from the incoherent part of the single-electron spectral
weight, as DMFT calculations have shown~\cite{LiuNatPhys12,Park11,Toschi}, but is very difficult
to obtain
from the single-electron excitations in the vicinity of the small Fermi pockets in the
weak-coupling
approaches. Indeed, an RPA calculation fails to represent the data \cite{LiuNatPhys12}.

The observation of the Mott insulating behavior in the iron oxychalcogenides
\cite{Zhu10}
and vacancy-ordered  alkaline iron selenides \cite{MFang,DMWang}
has provided further evidence for the strong electron correlations.
These materials should be
metallic according to the LDA bandstructure calculations in the paramagnetic
state (even in the presence of the ordered vacancies in the alkaline iron pnictides), but are
observed to be insulating with a very large ordered moment.
As discussed in Secs.~\ref{mott_oxyselenides} and \ref{kfese},
the Mott insulating nature
of these materials - as opposed to the metallic behavior of the
parent iron pnictides -
can be understood in terms of a reduced bandwidth
but a comparable strength of the Coulomb interaction
(along with the Hund's coupling). Since the reduction of the LDA bandwidth is rather modest,
the observed Mott insulating behavior provides evidence for the strong electron correlations
in the iron pnictides as well.
More generally, the evolution in the degree of itinerancy across the various materials families
in the
iron-based materials
can be understood in a picture in which the  strength of the local interactions ($U$)
is fixed but the kinetic energy($\propto W$) is variable, as can be seen through calculations
of multi-orbital models
\cite{Yu_multi1,Yu_multi2}  and {\it ab initio} LDA+DMFT studies \cite{Yin}.

\begin{figure}[t!]
\centering
\includegraphics[width=28pc]{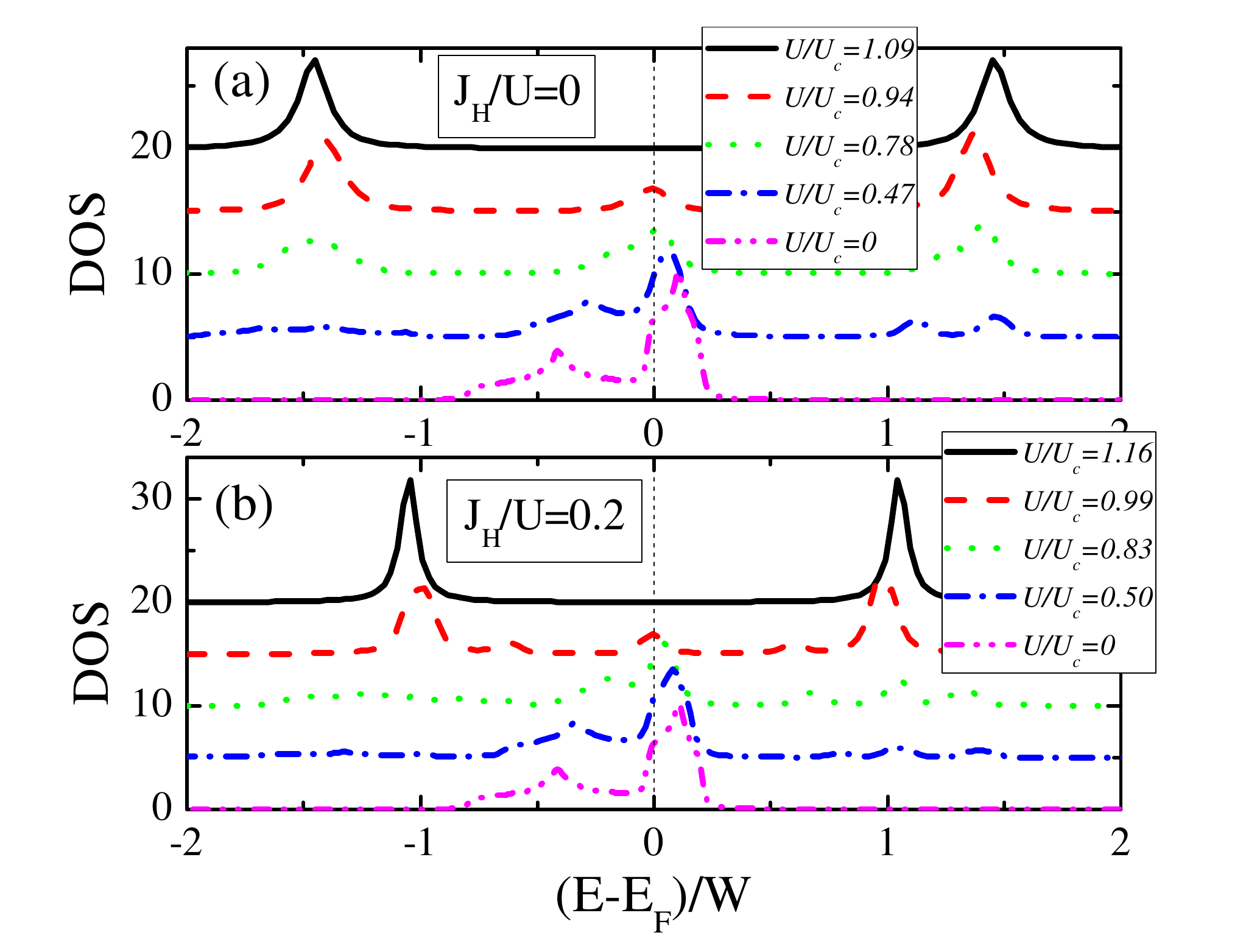}
\caption{\label{fig:dosslavespin}
{\bf Density of states (DOS) of a two-orbital model derived in a slave-spin study.}
Results for both the zero Hund's coupling (a) and a non-zero Hund's coupling (b) 
show incoherent electronic excitations across the Mott transition, $U=U_c$.
In both (a) and (b), the curves for $U/U_c>0$ have been shifted upward for clarity. 
Adapted from Ref.~\cite{Yu_multi1}.
}
\end{figure}

\section{The $w$-expansion of a multi-orbital Hubbard model with respect to the Mott transition}
\label{w_expansion}

For bad metals near a Mott transition, the incoherent electronic excitations play a dominant role
in the spin dynamics. This effect is difficult to describe in a direct expansion of $U/W$,
with respect to the non-interacting reference point. It can instead be more readily captured
by an expansion with respect to the Mott transition point,
with $(U_c-U)/W$ being the small parameter. In practice, this is equivalent to expanding in terms of $w$,
the percentage of the single-electron spectral weight in the coherent
part \cite{SiAbrahams,SiNJP,Moeller94}.

When $w=0$, at the Mott transition, we can
integrate out all the gapped single-electron excitations.
The vanishing $w$ implies the absence of the coherent spectral weight, as is already inferred in Fig.\ 1
and is further illustrated in Fig.~\ref{fig:dosslavespin}, which shows
the spectral function 
across the Mott transition $U_c$ in a multi-orbital model with or without the Hund's coupling $J_H$
as determined using a slave-spin technique ~\cite{Yu_multi1,Yu_multi2}.
This leads to localized magnetic moments, with the exchange interactions on the Fe-square lattice
containing both the nearest-neighbor component $J_1$ and the next-nearest-neighbor one $J_2$
\cite{SiAbrahams,SiNJP}.
General considerations suggest that $J_2 \gtrsim J_1/2$. In this parameter regime, the ground
state of the $J_1-J_2$ Heisenberg model on the square lattice is expected to have the collinear
$(\pi,0)$ antiferromagnetic order \cite{Larkin}, as is observed experimentally in the iron pnictides. Furthermore,
the antiferromagnetic order in this model is accompanied by an Ising order, whose linear coupling
to structural degrees of freedom induces a structural distortion which is also seen experimentally in the pnictides.

Because the charge gap is relatively small, we may also expect significant higher-spin couplings.
Of particular interest is the four-spin biquadratic interaction $K$, whose effect we will discuss
in Sec.~\ref{spin_dynamics}.

At higher order in $w$, there are itinerant coherent carriers which are coupled to the local moments.
This is summarized in the following equations for the couplings:
At $w=0$,
\begin{equation}
H_J  =
\sum_{\langle ij\rangle} J_1^{\alpha\beta}
{\bf s}_{i,\alpha} \cdot {\bf s}_{j,\beta}
+
\sum_{\langle\langle ij\rangle\rangle} J_2^{\alpha\beta}
{\bf s}_{i,\alpha} \cdot {\bf s}_{j,\beta}
+
\sum_{i,\alpha \ne \beta}
J_H ^{\alpha\beta}
{\bf s}_{i,\alpha} \cdot {\bf s}_{i,\beta},
\label{H_J}
\end{equation}
where $\langle ij\rangle$ and
$\langle\langle ij\rangle\rangle$ label
nearest-neighbor (n.n.) and next-nearest-neighbor (n.n.n.)
Fe sites on its square lattice. The greek indices label the orbitals and
 $J_H$ is
the on-site Hund's coupling.
Because of the multiple orbitals, both $J_1$ and $J_2$ are matrices.
At linear order in $w$, the kinetic energy $H_c$ of the itinerant carriers and their coupling
$H_m$ to the quasi local moments appear when integrating out the high energy degrees of freedom:

\begin{eqnarray}
&H_c&= w \sum_{{\bf k},\alpha,\sigma} E_{{\bf k}\alpha\sigma}
c_{{\bf k} \alpha \sigma}^{\dagger}
c_{{\bf k} \alpha \sigma}\\
&H_m& = w \sum_{{\bf k} {\bf q} \alpha\beta\gamma}
G_{{\bf k},{\bf q}\alpha\beta\gamma}~
c_{{\bf k}+{\bf q}\alpha\sigma}^{\dagger}
\frac{\btau_{\sigma\sigma^{\prime}}}{2}
c_{{\bf k}\beta\sigma^{\prime}} \cdot {\bf s}_{\mathbf{q}\gamma},
\end{eqnarray}
where $\btau$ are the Pauli matrices.
The final form for the low-energy effective Hamiltonian is
\begin{eqnarray}
H_{eff} = H_J + H_c + H_m .
\label{Heff}
\end{eqnarray}


\section{Quantum criticality in the iron pnictides}\label{qcp}
An important consequence of the strong-coupling approach is that by tuning the strong-coupling
parameter $w$, one can access a QCP, separating a paramagnetic metallic phase from
an antiferromagnetic or possibly insulating phase. Indeed, it was predicted \cite{DaiPNAS}
and subsequently
experimentally confirmed that doping phosphorus into a magnetic Fe-As parent compound would
reveal a magnetic quantum critical point, as shown in Fig.\ 4.
The reason is that, since P is smaller than As, the lattice contracts a bit upon P doping,
with a consequence that the kinetic energy (bandwidth) increases so that $w$ increases
to a critical value $w_c$ at which the magnetism disappears. This can be understood by examining
the action that describes the quantum fluctuations of the $J_1$-$J_2$ model, which is appropriate
when $w=0$: ${\cal S} = {\cal S}_2 + {\cal S}_4 + \cdot \cdot \cdot $, with

\begin{equation}
{\cal S} = \int d {\bf q} d\omega[r_0 +c_1q_x^2+
c_2 q_y^2 + \omega^2 [|{\bf m}({\bf q},\omega)|^2 + |{\bf m}'({\bf q},\omega)|^2].
\end{equation}
Here, $r_0<0$ and ${\bf m}$ and ${\bf m}'$ are vector order parameters describing
the N\'eel vectors of two interpenetrating sublattices on the square lattice
[with ${\bf m} \pm {\bf m}'$ respectively corresponding to the
($\pi,0)$ and $(0,\pi)$ order].
The couplings to the itinerant fermions at finite order in $w$ bring about
added terms to this effective theory.
To the linear order in $w$, there are two added terms:
\begin{equation}\label{DAMPING}
\Delta{\cal S} = \int d {\bf q} d\omega[ wA_{\bf Q} +\gamma|\omega| [|{\bf m}({\bf q},\omega)|^2
+ |{\bf m}'({\bf q},\omega)|^2]  
\end{equation}

The $wA_Q$ term is positive, providing a mass shift that weakens the $(\pi, 0)$ order;
when $r_0 +wA_Q =0$, the magnetic order disappears at a QCP.
The $\gamma|\omega|$ term describes the Landau damping; here the prefactor $\gamma$
is order $w^0$, but the linear-in-$\omega$ dependence
has an upper cutoff frequency that is linear in $w$.
Thus, the total ${\cal S} + \Delta{\cal S}$ gives rise to a magnetic QCP, as illustrated in Fig. 4,
where at a critical value of the P-doping, $r_0 +wA_Q =0$. 
An important ingredient of the analysis given in Ref.~\cite{DaiPNAS} is that the quartic coupling,
$\tilde{u} ({\bf m}\cdot{\bf m}')^2$, is only marginally relevant with respect to the underlying fixed 
point associated with the O(3) transition of ${\bf m}$ (and ${\bf m}'$). This makes the Ising and
antiferromagnetic transitions to be essentially coinciding and second order.
(By contrast, for the thermally driven transition, $\tilde{u}$ is strongly relevant
with respect to the classical O(3) fixed point, and the Ising and antiferromagnetic transitions 
become either significantly split or first order.)

This theoretical proposal for a magnetic QCP in P-doped parent iron arsenides
\cite{DaiPNAS}
has now been extensively confirmed in experiments carried out
in P-doped CeFeAsO \cite{delaCruz,Luo,Jesche} and BaFe2As2 \cite{Jiang09,Kasahara10,Hashimoto12}.

\begin{figure}[t!]
\centering
\includegraphics[totalheight=0.35\textheight, viewport=-50 100 800 500,clip]{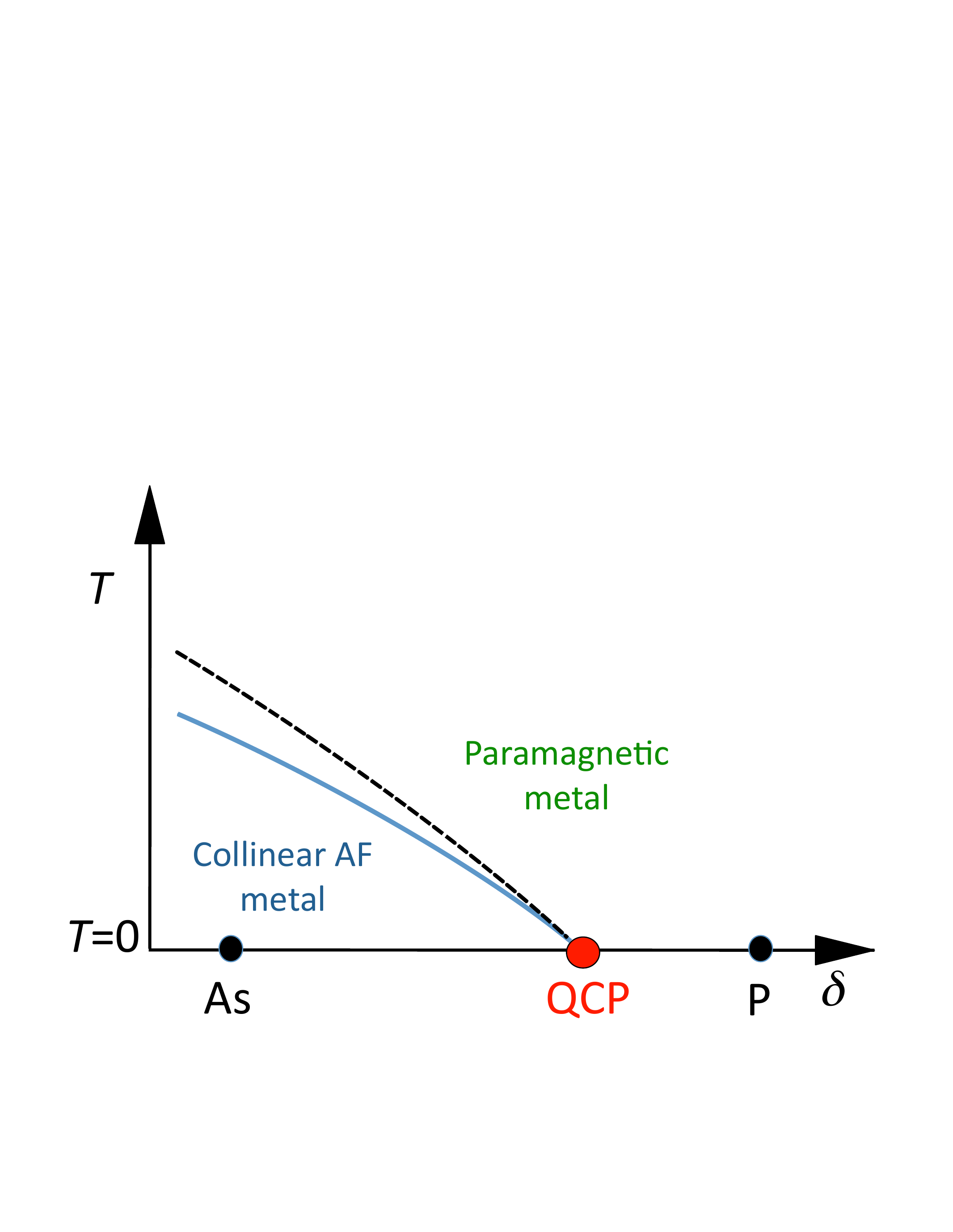}
\vskip -2pc
\caption{\label{fig5}
\textbf{Pnictide phase diagram near a magnetic quantum critical point.} Here $\delta$
 is a non-thermal control parameter: increasing $\delta$ enhances the spectral weight
 $w$ in the coherent part of the single-electron excitations. A specific example for $\delta$
 is the concentration of P doping for As. The red dot denotes the quantum critical point that
 separates a two-sublattice
collinear AF ground state from a paramagnetic one. The blue solid/black dashed lines respectively
represent the magnetic/structural transitions at non-zero temperatures.
Here the structural transition reflects the transition associated with the Ising order parameter,
${\bf m}\cdot{\bf m}'$.
Adapted from Ref.~\cite{DaiPNAS}.
}
\end{figure}

\section{Spin dynamics in the paramagnetic phase of parent iron pnictides}
\label{spin_dynamics}

Quantum criticality represents one way to study the quantum fluctuations, through the tuning
of a control parameter.  Alternatively, these fluctuations can be directly probed by studying
the frequency and wave-vector dependences of the spin structure factor. A particularly instructive
case to consider is the parent arsenides in their paramagnetic phases,
at $T>T_N$ \cite{Diallo10,Harriger11,Ewings11}; the absence of order fascilitates the study
of the underlying interactions. With that in mind, we have studied the
$J_1-J_2-K$ model:

\begin{equation}
H= J_1\sum_{i,\delta}{\bf S}_i\cdot{\bf S}_{i,\delta}
+ J_2\sum_{i,\delta'}{\bf S}_i\cdot{\bf S}_{i+\delta'}
- K\sum_{i,\delta}({\bf S}_i\cdot{\bf S}_{i,\delta})^2
\end{equation}
which is combined with the damping term specified in Eq.\ \ref{DAMPING}.
Our approach represents one way
to describe the spin dynamics by incorporating the contributions of the incoherent single-electron
excitations
in bad metals close to a Mott transition. An alternative approach is to study the spin dynamics
in this regime
using the DMFT method \cite{Park11}. The results from the two approaches are complementary.

\begin{figure}[t!]
\centering
\includegraphics[totalheight=0.4\textheight
,clip]{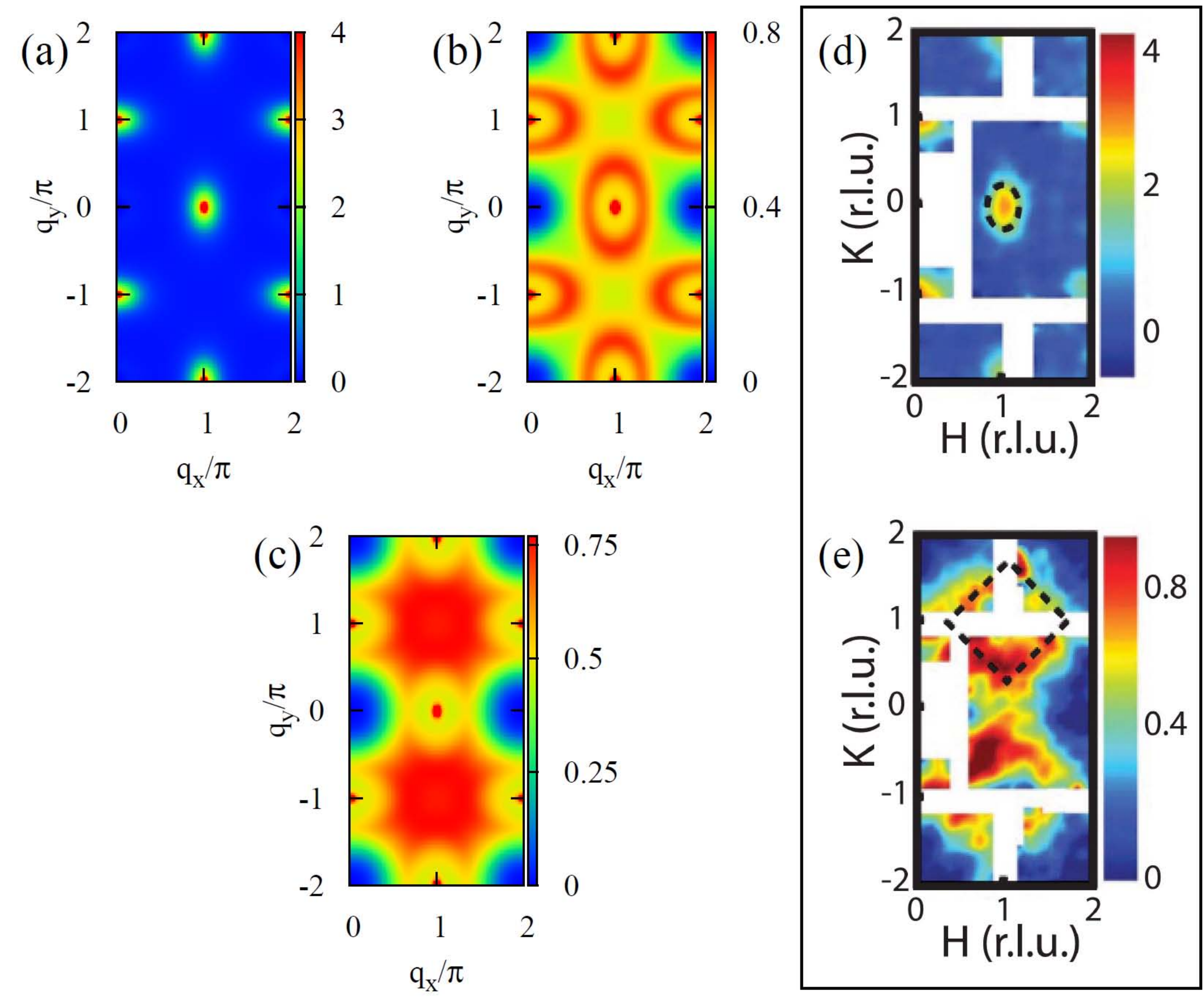}
\caption{\label{spindynamicsJ1J2K}
\textbf{Evolution of the spin structure factor $S(q,\omega)$ in the paramagnetic phase
of the $J_1-J_2-K$ model and the comparison with inelastic neutron scattering results
of BaFe$_2$As$_2$.} The calculated $S(q,\omega)$ shows elliptical features near $(\pi,0)$
at low energies (in panels (a) and (b)). They are split into features that are centered around $(\pi,\pi)$,
as the energy is increased towards the zone-boundary
spin-excitation energy (panel (c)). This trend is consistent with the inelastic neutron scattering
experiments, shown in the box at the right for two energies measured in the paramagnetic phase
of BaFe2As2 (data taken from Ref.~\cite{Harriger11}). Adapted from Ref.~\cite{YuJ1J2K}.
}
\end{figure}

To calculate the dynamical structure factor, in Ref. \cite{YuJ1J2K} we have
decoupled the biquadratic coupling using
Hubbard-Stratonovich fields and applied a modified spin-wave method \cite{Takahashi}
to treat the resulting
spin Hamiltonian. At low energies, $S({\bf q},\omega)$ as a function of ${\bf q}$
displays elliptic features centered
around $(\pi,0)$ and its symmetry-equivalent points. Such elliptic features reflect the existence
of two correlation lengths whose ratio is controlled by $J_2/J_1$, and are not
sensitive to the magnitude of the biquadratic  coupling $K$.
[Indeed, the results at low energies are similar to those for the $K=0$ case
\cite{Goswami_J1J2}.]
At high energies, $S({\bf q},\omega)$ vs. ${\bf q}$ is sensitive to the $K$ coupling. Fig.~\ref{spindynamicsJ1J2K} shows the results
of the energy evolution of the momentum-distribution of the dynamical structural factor,
which compares well
with the experimental observations \cite{Harriger11}.

For the insulating iron alkaline iron selenides in the presence of the $\sqrt{5}\times \sqrt{5}$
vacancy order, both the block spin order and spin-wave excitations can be studied
in an extended
$J_1-J_2$ model~\cite{YuGoswami11}.
Spin-wave measurements have fully confirmed the calculated
spectrum~\cite{WangDai11}.

\begin{figure}[t!]
\centering
\includegraphics[totalheight=0.5\textheight, viewport=-50 -10 800 450,clip]{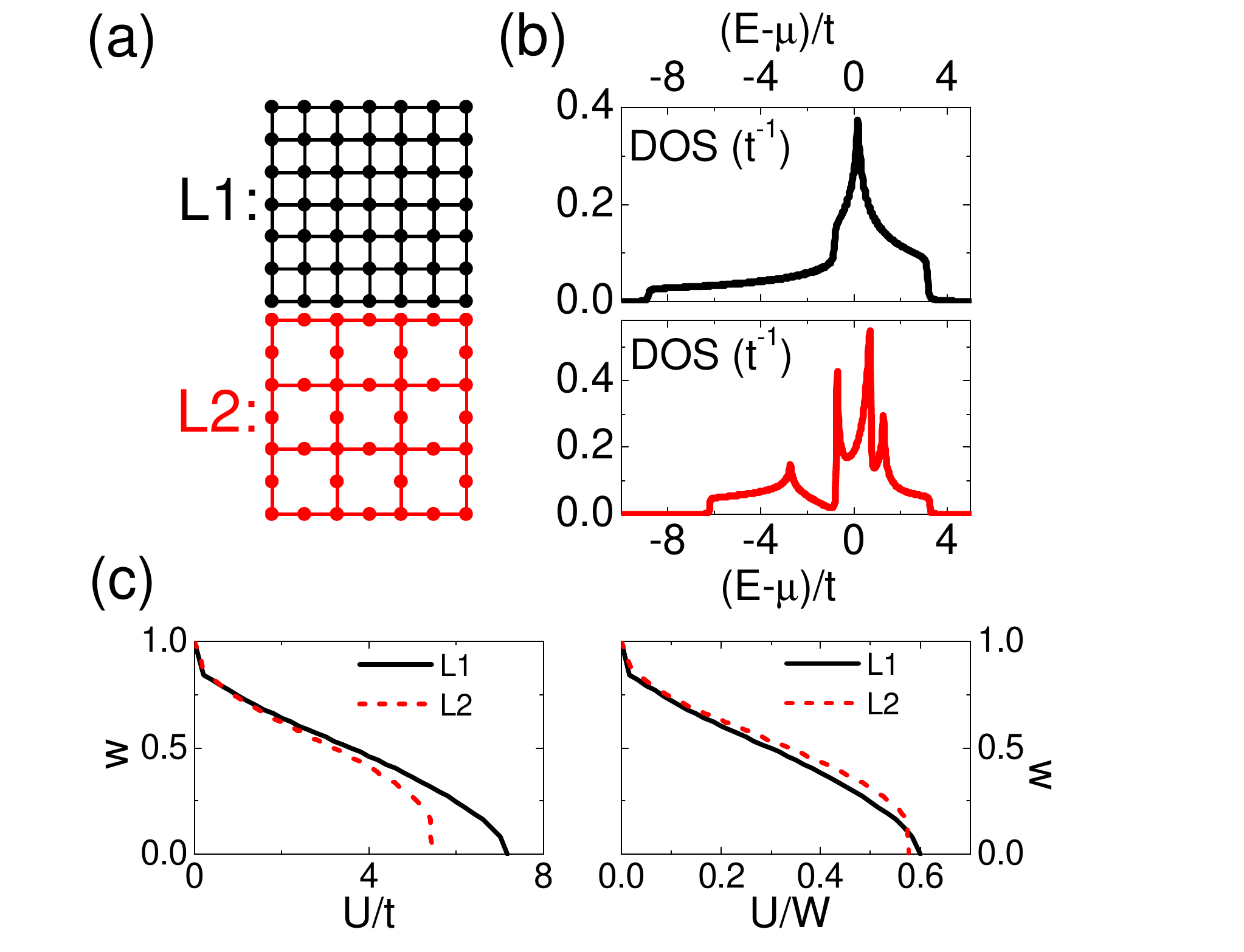}
\vskip -1pc
\caption{\label{mott_due_to_vacancy}
\textbf{Kinetic energy reduction and Mott localization by ordered Fe vacancies.}
(a) The regular square lattice and  its counterpart with $2 \times 2$ ordered vacancies
(the pattern is chosen for an illustrative purpose, such that the model calculation is simplified,
and similar results arise
 for other patterns including $\sqrt{5}\times \sqrt{5}$);
(b) DOS of the two lattices;
(c) Left panel: $w$ vs. $U/t$. The Mott threshold $U_c$ is reduced by the ordered vacancies.
Here, $t$ is the nearest-neighbor hopping parameter.
Right panel: $w$ plotted against $U/W$,
showing that
$U_c$  tracks  $W$, the bandwidth.
Adapted from Ref.~\cite{Yu_prl11}.
}
\end{figure}

\section{Parent Mott insulator arising from kinetic energy reduction}
\label{mott_oxyselenides}
In Sec.~\ref{qcp}, we discussed how to induce a QCP by increasing the kinetic energy,
which increases $w$ and weakens the magnetic order. In the same spirit, one would
expect
to be able to tune the system in the opposite way, by reducing the kinetic energy and
thereby increasing the correlation effect.
This corresponds to moving to the left on the $U/t$ axis shown in Fig.~\ref{QMCap}.

The iron oxychalcogenides
La$_2$O$_2$Fe$_2$O(Se,S)$_2$ provide a case study \cite{Zhu10}.
These  are also
parent compounds, {\it i.e.} they have a composition such that  the nominal valence of
Fe is $2+$.
They also contain an Fe square lattice, which is expanded compared to that
of the iron pnictides; this leads to narrower Fe $3d$-electron bands.
According to LDA calculations, the paramagnetic ground state should be metallic.
However, the band narrowing enhances correlation effects
and promotes the Mott insulating state.
Measurements of the transport and magnetic properties provide the experimental evidence
that the system is indeed a Mott insulator.  Because the bandwidth
of La$_2$O$_2$Fe$_2$O(Se,S)$_2$ is narrower than that of the parent
iron pnictides by only a moderate amount (on the order of 25\%),
this result provides evidence that
the parent pnictides are on the verge of Mott localization. Furthermore, elastic neutron scattering 
measurements have determined the magnetic order \cite{lofese_neutron}.
and found a large ordered moment, about $2.8$ $\mu_B$/Fe, consistent with the discussion
in Sec.\ \ref{bad_metal}.

Mott insulating behavior also exists in
 R$_2$O$_2$Fe$_2$OSe$_2$ (R=Ce,Pr,Nd, and Sm) which possess
a similarly
 large ordered moment \cite{Ni}
and in several other iron oxychalcogenides whose ordered moments are yet to be determined:
 (Sr,Ba)$_2$F$_2$Fe$_2$O(Se,S)$_2$ (Ref.~\cite{Kabbour}),
 Na$_2$Fe$_2$OSe$_2$ (Ref.~\cite{JBHe}),
and
BaFe2OSe2 (Refs.~\cite{Han,Lei}).

\begin{figure}[t!]
\centering
\includegraphics[totalheight=0.4\textheight, viewport=-100 0 700 450,clip]{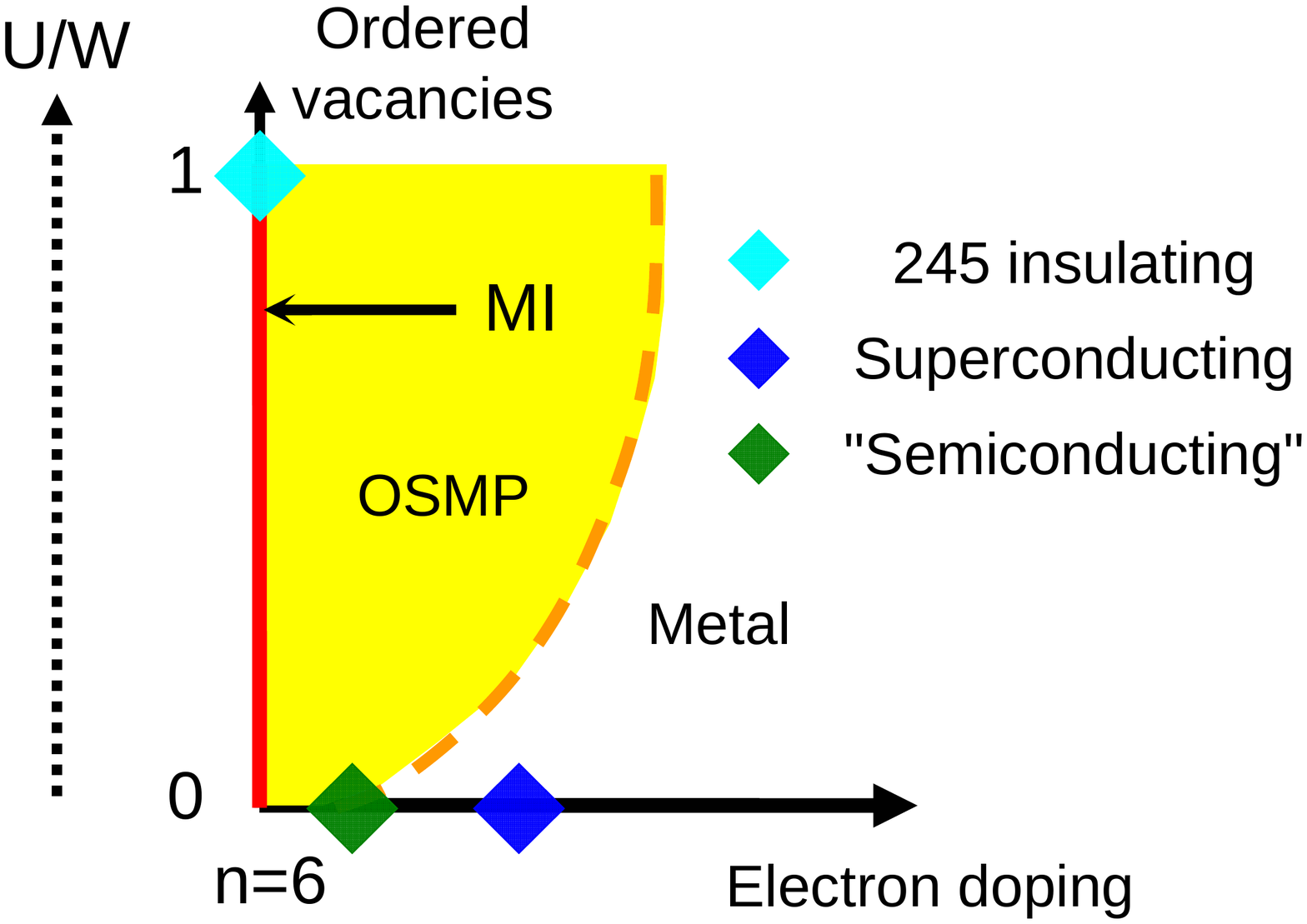}
\caption{\label{osmp}
\textbf{Overall phase diagram for {\rm K$_{1-x}$Fe$_{2-y}$Se$_2$}.}
The vertical axis represents the degree of iron vacancy order,
with 1 corresponding to fully $\sqrt{5}\times\sqrt{5}$ vacancy ordering and 0
marking
complete vacancy disorder or absence of vacancies.
The degree of the iron vacancy order tunes  $U/W$.
MI  and OSMP stand for the Mott insulator
and the orbital-selective Mott phase, respectively,
and the orange dashed line describes the transition from OSMP to the delocalized
metallic phase.
The diamond symbols mark the approximate positions for the various
phases that have been observed experimentally.
}
\end{figure}

\section{Alkaline iron selenides: Mott insulator and orbital-selective Mott insulator}
\label{kfese}
Recently, superconductivity with $T_c$ comparable to the iron pnictides was
discovered in K$_{\rm {1-y}}$Fe$_{2-x}$Se$_{2}$ (``KFS")
\cite{Guo}
and
related iron selenides with $\mathrm{K}$ replaced
by
$\mathrm{Rb}$, $\mathrm{Cs}$, or  $\mathrm{Tl}$ \cite{MFang,Krzton-Maziopa,Mizuguchi}.
Here too,  the superconductivity occurs near  antiferromagnetic order~\cite{MFang}.
Angle-resolved photoemission (ARPES) experiments \cite{Zhang_Feng,Qian_Ding,Mou}
have shown that the Fermi surface has only electron pockets, in contrast to the iron pnictides ,
which have both electron and hole pockets.

The parent selenide compounds, with Fe valence at $2+$, are insulating~\cite{MFang,Wang1101_0789}.
These compounds have Fe vacancies forming ordered patterns
~\cite{MFang, Bao245}. LDA band-structure calculations show that the paramagnetic ground
state of such a vacancy-ordered structure is metallic \cite{CaoDaiPRL11,Yan11},
therefore the insulating ground state must result from electron interactions. How strong are these
interactions?
A clue is provided by the size of the ordered moment, which is about 3.3 $\mu_B$/Fe
\cite{Bao245,WangDai11}.
Recognizing that
the maximal possible moment is 4 $\mu_B$/Fe, the observed ordered moment is indeed very
large. Therefore, the interaction energy must be sizable compared to the kinetic energy.

Microscopic considerations have suggested that the Mott insulating nature of the vacancy-ordered
system is very natural \cite{Yu_prl11}.
The ordered vacancies lead to a band-narrowing,
which enhances the ratio $U/W$, and pushes
the system to the Mott insulator side. This is shown in Fig.~\ref{mott_due_to_vacancy},
with an illustrative model having a
 $2 \times 2$ vacancy ordering pattern~\cite{Yu_prl11}.
 
In these iron-based compounds, there are five 3$d$ orbitals which are non-degenerate; a slave-spin
study \cite{YuSi12}
has identified an orbital-selective Mott phase (``OSMP") in which the $xy$ orbitals are localized
and the remaining 3$d$ orbitals are delocalized. ARPES measurements in the superconducting
compound have shown that raising the temperature
causes a suppression of the spectral weight of the $xy$ orbitals near the Fermi energy,
while keeping the weight of its $xz$ and $yz$ counterparts non-zero
\cite{Yietal12}, Combining the ARPES results and the theoretical phase diagram yields
an overall phase diagram
shown in Fig.~\ref{osmp}.

The OSMP provides the link between the vacancy-ordered/Mott insulating phase and
the vacancy disordered/free superconducting phase. Experimental evidence has come from
transport measurements in the insulating alkaline
iron selenides under pressure  \cite{PGao}.

We close this section by discussing the expected paramagnetic spin excitations
in the insulating Rb$_{0.8}$Fe$_{1.6}$Se$_2$.
Because $T_N$ is very large (on the order of 500 K), and is only slightly (a few 10's K)
below the vacancy
ordering temperature $T_s$, 
the $\sqrt{5}\times \sqrt{5}$ iron vacancy order becomes fragile above $T_N$~\cite{Bao245}.
As we have just discussed in this section, the reduction of the iron vacancy order
leads to an effective reduction of the electron correlation strength $U/W$ from  that of the 
fully vacancy-ordered case,
destabilizing the Mott insulating state. Correspondingly, the spin spectral weight will be reduced
from the very large value in the fully vacancy-ordered state.
This has recently been observed:
Above the N\'{e}el temperature, Wang et al.~\cite{WangDai2012} reported spin excitations that are 
more strongly damped than those of the fully vacancy-ordered state as well as  
a significant reduction of the total moment. 
In other words, the experimental observations in the paramagnetic regime, both below and above the vacancy ordering
temperature, are consistent with the Mott-insulating behavior of the fully vacancy-ordered state.
The fact that the magnetic excitations at low temperatures are well described by spin waves
with a very large stiffness constant reinforces the Mott-insulating behavior of the fully vacancy-ordered state.

 \begin{figure}[t!]
\centering
\includegraphics[totalheight=.4\textheight, viewport=0 0 600 160,clip]{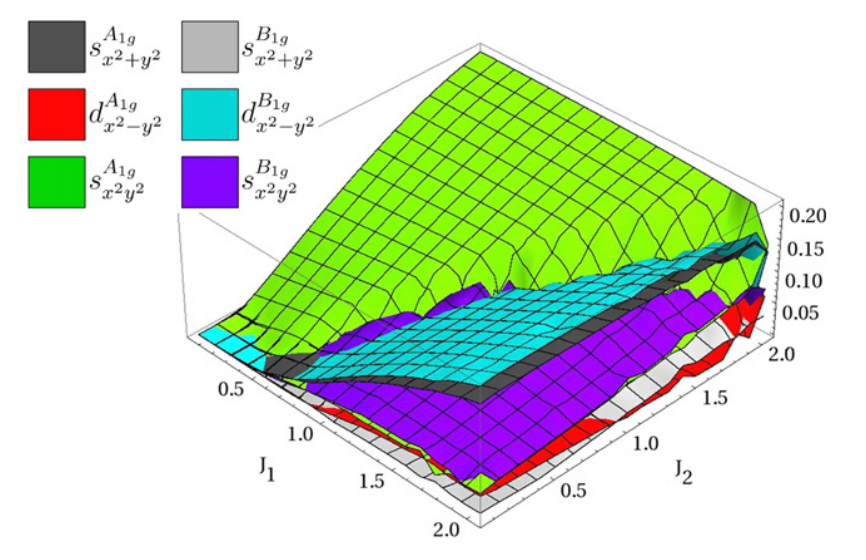}
\caption{\label{sc_pnictides}
\textbf{The pairing amplitudes (for $d_{xz}$, $d_{yz}$ orbitals) in the
five orbital model for electron doping $\delta=0.14$.}  Adapted from Ref.~\cite{Goswami}.}
\end{figure}

\section{Superconductivity in iron pnictides and alkaline iron selenides}
\label{sc}

In the carrier doped case, the $w$-expansion discussed earlier leads to a 5-band
$t-J_1-J_2$ model \cite{SiNJP,SiAbrahams}.
Spin-singlet superconducting pairing from this
model was studied in Ref.~\cite{Goswami} in the case of the iron pnictides.
Magnetic frustration by itself leads to a large degeneracy in the pairing
states. The kinetic energy breaks this into a quasi-degeneracy among
a reduced set of pairing states,
as shown in Fig.~\ref{sc_pnictides}.
For small
electron and hole Fermi pockets,
an $\mathrm{A}_{1g}$ $\mathrm{s}_{x^2y^2}$
state dominates over the phase diagram
but a $\mathrm{B}_{1g}$ $\mathrm{d}_{x^2-y^2}$
 state is close by in energy; an $\mathrm{A}_{1g}+i \mathrm{B}_{1g}$ state,
which breaks time-reversal symmetry, occurs at
low temperatures in part of the phase diagram.
Compared to the two-band case, there is an enhanced amplitude for the
 $\mathrm{A}_{1g}$ $\mathrm{s}_{x^2+y^2}$ channel, which causes anisotropy of the
 gap on the electron pockets or even an accidental gap near $M$.

 An important advantage of this strong-coupling approach is that it gives rise to
 pairing amplitudes for the alkaline iron selenides that are comparable to those for the iron
 pnictides, in spite of the very different Fermi surfaces. This is consistent with the fact that
 the superconducting transition temperatures are very similar.

\section{Summary and outlook}
\label{summary}

We have summarized the phenomenological basis for strong electron correlations in the iron
pnictides
and chalcogenides. Chief among these are the bad-metal
behavior seen in the electrical transport and charge dynamics properties in both classes of materials;
the large spectral
weight observed through spin dynamics measurements, also seen in both cases;
and the observation of Mott insulating states in the iron oxychalcogenides and alkaline iron selenides.

We have also outlined a $w$-expansion, which uses the Mott transition as a reference point.
This expansion treats the bad-metal behavior through a proximity to the Mott
transition, which separates the single-electron excitations into coherent
and incoherent ones whose weights are $w$ and $1-w$ respectively.
In this way, it naturally incorporates the large spin spectral weight
in the bad-metal ({\it i.e.}, small $w$) regime.

For the parent arsenides, this expansion gives rise to an effective spin Hamiltonian in the form of
a $J_1$-$J_2$ model that is coupled to the itinerant electrons in the vicinity of the Fermi energy.
Varying $w$ amounts to tuning the relative degree of itinerancy vs. localization and provides a means
to vary the strength of magnetic ordering. We have discussed an early theoretical proposal for
a quantum critical point resulting from $w$-tuning and for its realization through a P-for-As
isoelectronic substitution in the parent iron arsenides. This prediction has since been
experimentally confirmed by a number of groups using transport, neutron, NMR
and quantum oscillation measurements. Using the same approach,
theoretical studies have been carried out for the spin dynamics  in the paramagnetic phase
of the parent arsenides.  In the presence of a biquadratic interactions, the theoretical
results compare well with the inelastic neutron scattering results above $T_N$,
both at low energies near the magnetic-zone centers
and at high energies close to the magnetic-zone boundaries.

We have also demonstrated how a Mott insulating phase
 may arise from a kinetic-energy reduction, and its experimental evidence
 in both the iron oxychalcogenide compounds and the vacancy-ordered alkaline iron selenides.
 In the latter families, we show how an orbital-selective Mott phase appears in the phase diagram.

Finally, we have discussed the singlet superconducting pairing driven
by the short-range $J_1$-$J_2$ interactions.
For the carrier-doped cases, the $w$-expansion gives rise to
a five-orbital
matrix $t$-$J_1$-$J_2$ model. In this model,
the dominant pairing channels are the nearly degenerate
 $\mathrm{A}_{1g}$ $\mathrm{s}_{x^2y^2}$
and $\mathrm{B}_{1g}$ $\mathrm{d}_{x^2-y^2}$
 states, with $\mathrm{A}_{1g}$ $\mathrm{s}_{x^2+y^2}$ channel also playing a significant
 role in the iron pnictides.

More generally, we have highlighted the phenomenological and theoretical evidences that
the iron pnictides and chalcogenides are bad metals at the boundary between electronic
localization and itinerancy. As such, these materials not only are important in their own right,
but also provide a setting to study the physics in this important parameter regime of strongly
correlated
electrons in general.

\section*{Acknowledgments}

We are grateful to J. Dai, A. H. Nevidomskyy, P. Nikolic,  Z. Wang and J.-X. Zhu
for collaborations, and
to them as well as many other colleagues for useful discussions. This work has been in part
supported
by the NSF Grant No. DMR-1006985 and the Robert A. Welch Foundation Grant No. C-1411.
P. G. was supported at the National High Magnetic Field Laboratory by NSF Cooperative
Agreement No. DMR-0654118, the State of Florida, and the U. S. Department of Energy.
Part of this work was carried out at the Aspen Center for Physics (NSF grant 1066293).

\smallskip
\end{document}